\documentclass[10pt,authoryear]{elsarticle}
\usepackage{amsmath,amssymb,amsthm,mathrsfs,amsopn}
\usepackage{graphicx}
\usepackage{array}
\usepackage{subfigure}
\usepackage{booktabs}

\journal{The Journal of Neuroscience Methods}

\begin{document}

\begin{frontmatter}

\title{\bf Improving Network Inference: 
The Impact of False Positive and False Negative Conclusions about the Presence or Absence of Links}

\author[1,2]{Gloria Cecchini\corref{cor1}}\ead{$gloria.cecchini@abdn.ac.uk$}
\author[1]{Marco Thiel}\ead{$m.thiel@abdn.ac.uk$}
\author[1]{Bj\"orn Schelter}\ead{$b.schelter@abdn.ac.uk$}
\author[1]{Linda Sommerlade}\ead{$l.sommerlade@abdn.ac.uk$}
\cortext[cor1]{Corresponding author \\ Address: room 343, Meston Building, Meston Walk, Aberdeen, AB24 3UE, United Kingdom \\ Telephone Number: +441224272317}
\address[1]{Institute for Complex Systems and Mathematical Biology, University of Aberdeen, Meston Building, Meston Walk, Aberdeen, AB24 3UE, United Kingdom}
\address[2]{Institute of Physics and Astronomy, University of Potsdam, Campus Golm, Karl-Liebknecht-Stra{\ss}e 24/25, 14476, Potsdam-Golm, Germany}

\begin{abstract}

\subsection*{Background}
A reliable inference of networks from data is of key interest in the Neurosciences.
Several methods have been suggested in the literature to reliably determine links in a network.
To decide about the presence of links, these techniques rely on statistical inference, typically controlling the number of false positives, paying little attention to false negatives.
\subsection*{New Method}
In this paper, by means of a comprehensive simulation study, we analyse the influence of false positive and false negative conclusions about the presence or absence of links in a network on the network topology. 
We show that different values to balance false positive and false negative conclusions about links should be used in order to reliably estimate network characteristics. 
We propose to run careful simulation studies prior to making potentially erroneous conclusion about the network topology.
\subsection*{Results}
Our analysis shows that optimal values to balance false positive and false negative conclusions about links depend on the network topology and characteristic of interest.
\subsection*{Comparison with Existing Methods}
Existing methods rely on a choice of the rate for false positive conclusions. 
They aim to be sure about individual links rather than the entire network. 
The rate of false negative conclusions is typically not investigated.
\subsection*{Conclusions}
Our investigation shows that the balance of false positive and false negative conclusions about links in a network has to be tuned for any network topology that is to be estimated.
Moreover, within the same network topology, the results are qualitatively the same for each network characteristic, but the actual values leading to reliable estimates of the characteristics are different.
\end{abstract}

\begin{keyword}

network inference \sep node degree distribution  \sep false positive \sep false negative \sep statistical inference

\end{keyword}

\end{frontmatter}

\section{Introduction}

Recently, many research groups have focused on the inference of networks from data such as brain networks from observed electroencephalography or functional magnetic resonance imaging data \citep{Bullmore2009, Pessoa2014, Petersen2015, Sporns2004}. 
Particular emphasis is paid to the understanding of the normal functioning, e.g. healthy brain, as well as malfunctioning, e.g. diseased brain, of these networks.
In the example of the brain, this promises to disclose information about how the brain processes signals and how alterations thereof cause specific diseases. 
A key hypothesis is that important characteristics are not specific to individual subjects but rather common in a given population. 
This is reflected by the fact that brain networks, but also other networks, are typically classified into few main prototypic networks  \citep{NewmanBook, Newman2002}, e.g., Erd{\H o}s-R\'enyi \citep{Erdos1959,Erdos1960}, Watts-Strogatz \citep{watts1999small, Watts1998}, Barab{\'a}si-Albert \citep{barabasi1999,barabasi2016network} networks.
In our work we consider binary undirected networks of these three topologies.

These prototypical models for networks are in turn characterised by few parameters; procedures have been described to generate these networks with their well-established characteristics \citep{NewmanBook,Newman2002}.
Some of the key characteristics are the node degree distribution, the number of links, the global clustering coefficient, and the efficiency.
We considered these characteristics in our study since they are meaningful in random networks and give a global description in large networks \citep{NewmanBook}.

In the {\em Inverse Problem}, the challenge is to infer the network topology from data.
Two challenges are particularly relevant: (i) the reliable inference of links in the network once the nodes have been fixed \citep{Mader2015, zerenner} and (ii) the successful usage of the characteristics above to uniquely determine the topology of network \citep{Bialonski2010,Bialonski2011}.

The correct reconstruction of networks is hampered not only by false conclusions about links due to statistical uncertainties, but also by unobserved processes \citep{Elsegai2015, Guo2008, Ramb2013} and noise contamination \citep{Nalatore2007, Newbold, Sommerlade2015} to name just a few challenges of network reconstruction.
Classical statistical methods to estimate links in a network aim to identify present links with high certainty, e.g. \citep{jalili2011, quinn2002, devore2011, Schinkel2011, Fallani2014, Chavez2010, Honey2007}.
Therefore, typically the rate of false positive conclusions about links is chosen and consequences for the rate of false negative conclusions about links are accepted.
We investigate if these common rules of false positive conclusions and false negative conclusions should be modified to achieve a more reliable inference of the correct topology of network.
To this aim, we analyse their influence on the network topology and characteristic.

The manuscript is structured as follows.
An introduction to network topologies and their characteristics is given in Section \ref{NetworkCharacteristics}.
Section \ref{InferenceReliability} explains statistical errors and their influence on the network topology.
A simulation study in the case of Erd{\H o}s-R{\'e}nyi, Watts-Strogatz and Barab{\'a}si-Albert networks is presented in Section \ref{Results}.

\section{Materials and Methods}
\label{Materials and Methods}
In this section, network topologies and their characteristics are described (Section \ref{NetworkCharacteristics}).
We summarise statistical errors and suggest a distance measure to quantify their influence on the estimation of network characteristics (Section \ref{InferenceReliability}).

\subsection{Network Characteristics}
\label{NetworkCharacteristics}

A network $G$ is defined as a set of nodes with links between them. 
To quantify the topology of networks, different network characteristics have been described \citep{Olbrich2010}. 
Here, we consider four network characteristics: node degree, number of links, global clustering coefficient and efficiency. 
The node degree describes the number of links of a node.
For example, if the node $v$ has $k$ links attached, its node degree is $d_v=k$.
Typically the node degree distribution is used to characterise the entire network. 
The number of links refers to half of the sum over the node degrees.

The global clustering coefficient describes how well the neighbours of a node are connected. 
More precisely it measures the conditional probability that given one node connected to other two nodes, these are also connected to each other \citep{Olbrich2010}.

For two randomly selected nodes $i,j$ in a network of $n$ nodes, the shortest path length $\ell_{ij}$ measures the number of steps separating them if the shortest path is taken. 
The average path length $\gamma=\frac{1}{n(n-1)}\sum_{i\neq j}\ell_{ij}$ gives a measure of the sparsity of the network.
The efficiency $\epsilon=\frac{1}{n(n-1)}\sum_{i\neq j}\frac{1}{\ell_{ij}}$ is defined as the sum of the inverse of the shortest paths lengths.
Since the shortest path is infinitely long for unconnected nodes, taking the average of the shortest path length in a network with unconnected nodes is not meaningful.
Efficiency for unconnected nodes will be zero, therefore a meaningful network average of efficiency can be obtained.

Different network topologies have been described \citep{Newman2003}. 
Here, we investigate Erd{\H o}s-R{\'e}nyi \citep{Erdos1959, Erdos1960}, Watts-Strogatz \citep{watts1999small, Watts1998} and Barab{\'a}si-Albert \citep{barabasi1999} networks, as key examples of networks. 

Erd{\H o}s-R{\'e}nyi networks are random networks in which each pair of nodes is connected with independent probability $p_c$.
The probability mass function of the node degree distribution of a Erd{\H o}s-R{\'e}nyi network 
\begin{equation}
\mathbb{P}(d_v=k)=\binom{n-1}{k}p_c^k(1-p_c)^{n-1-k}
\end{equation}
is a binomial distribution \citep{NewmanBook}.

Watts-Strogatz networks are also referred to as small-world networks. 
They are characterised by a high local connectivity with some long-range ``short-cuts''. 
Watts-Strogatz networks are built from a regular network with node degree $2c$. 
Nodes are arranged on a circle; therefore, each node has $c$ nearest clockwise as well as $c$ nearest counterclockwise neighbours. 
With probability $p_r$ each link connecting a node to one of its nearest neighbours is reconnected to another node randomly chosen. 
The node degree distribution has probability mass function
\begin{equation}
\mathbb{P}(d_v=k)=\sum_{i=\max(2c-k,0)}^{\min(n-1-k,2c)}\binom{2c}{i}\left(\frac{p_r}{2}\right)^i \left(1-\frac{p_r}{2}\right)^{2c-i} e^{-cp_r}\frac{(cp_r)^{k-2c+i}}{(k-2c+i)!},
\end{equation}
in the assumption that $n\gg c$, \citep{Menezes2017}.
Here, we study Watts-Strogatz networks with $c=2$.

Barab{\'a}si-Albert networks are so-called scale free networks. 
They are constructed by adding nodes to an existing network.
The degree of the existing nodes influences the probability for a new link.
Each new node is connected to the network with a certain number $b$ of links.
The probability for one of these $b$ links to be formed with any existing node is proportional the degree of that node. 
The node degree distribution has probability mass function \citep{barabasi2016network}
\begin{equation}
\mathbb{P}(d_v=k)=\frac{2b(b+1)}{k(k+1)(k+2)}.
\label{eqBarabasi}
\end{equation}
Note that Eq.~(\ref{eqBarabasi}) is of type $\mathbb{P}(d_v=k)=c_1k^{-c_2}$, where $c_1$ and $c_2$ are constants, i.e., it follows a power law.

\subsection{Inference Reliability}
\label{InferenceReliability}

In the Neurosciences different methods to identify nodes of a brain network exist. 
For our purposes the method of identifying nodes is not relevant. 
Therefore, we assume a fixed set of nodes. 
Once nodes have been fixed, several methods have been suggested in the literature to address the challenge of reliable inference of links in the network. 
To determine the presence of links, these techniques usually rely on statistical inference, \citep{jalili2011, quinn2002, devore2011, Schinkel2011, Fallani2014, Chavez2010, Honey2007}.

Two types of errors exist when making these statistical inferences:
(i) an absent link ($\overline{C}$) may be erroneously assumed to be present by the method ($C^{D}$), this is a false positive conclusion and referred to as a \textit{type I error};
(ii) a present link ($C$) may remain undetected ($\overline{C^{D}}$) by the method, this is a false negative conclusion and referred to as a \textit{type II error}.
We call $\alpha=\mathbb{P}( C^{D} | \overline{C} )$ the probability of a false positive conclusion and $\beta=\mathbb{P}( \overline{C^{D}} | C )$ the probability of a false negative conclusion.
These two probabilities ($\alpha$ and $\beta$) are related and cannot be fixed independently. 
A standard choice is to set $\alpha=0.05$ and neglect investigation of $\beta$, focussing on reliably detecting individual links of the network.

Let $G$ denote the true network.
As a consequence of the choice of $\alpha$ and thereby $\beta$, leading to a non-zero probability of detecting false positive and false negative links, the network we detect $G^D$ will be a ``mixture'' of true links, false positive links, absent links and false negative links.
Therefore, the number of detected links is generally different to the number of links of $G$.
Also the node degree distribution, the global clustering coefficient and the efficiency are in general biased.
We quantify the bias for each characteristic using a distance between distributions. 
Several distance measures are conceivable and have been investigated; for sake of simplicity and to make the arguments clearer, we only consider the distance
\begin{equation}
\delta = |\mu_1 -\mu_2|
\label{dist}
\end{equation}
between two distributions, as the modulus of the difference of the distribution's mean values.
For example, the distance between the node degree distribution of $G$, which has mean $\mu_G$, and the node degree distribution of $G^D$, which has mean $\mu_{G^D}$, is $\delta = |\mu_G -\mu_{G^D}|$.

\begin{figure}[t]
\centering
\includegraphics[width=0.85\textwidth]{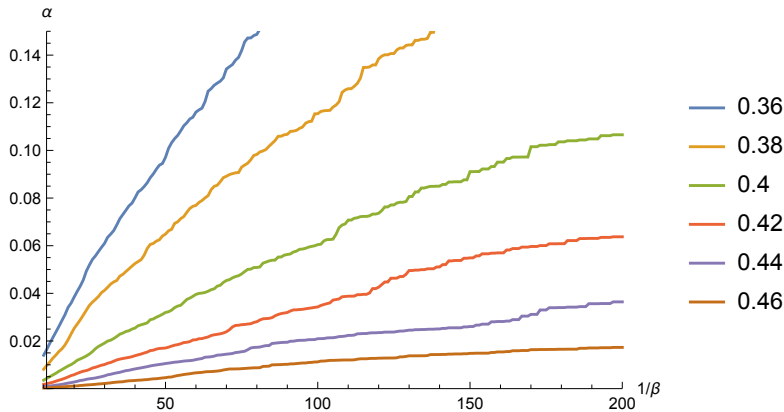}
\caption{Relation between $1/\beta$ and $\alpha$ for correlation of 100 simulated data points. Colours refer to different correlation coefficients used for the simulation, as indicated in the legend. The slopes of linear functions fitted to these curves vary between $0.1 \cdot 10^{-3}$ and $1.1 \cdot 10^{-3}$.}
\label{linear}
\end{figure}

To investigate the relation between $\alpha$ and $\beta$ numerically we simulate $N=100$ data points taken from a bivariate normal distribution.
To inspect in particular links with medium strength, we vary the simulated correlation between $0.36$ and $0.46$ in steps of $0.02$.
For each value, we repeated the simulation 10,000 times and tested for correlation using Pearson's correlation test.
From this test we inferred the probability that the 100 simulated data points are not correlated, the so-called $p-$value of the test.
Based on the 10,000 $p-$values of the correlation tests we inferred the relation between $1/\beta$ and $\alpha$ (see Fig.~\ref{linear}).
Visual inspection of Fig.~\ref{linear} shows that a linear relationship is a good approximation.
Fitting linear functions to the curves shows that their respective slopes vary between $0.1 \cdot 10^{-3}$ and $1.1 \cdot 10^{-3}$.
These slopes will differ if different simulation parameters, such as the number of data points N, are chosen.
The more data points are considered the more accurate the analysis.
Note that the inverse proportionality of $\alpha$ and $\beta$ implies that an infinite number of data points $N$ is needed to have both $\alpha$ and $\beta$ equal to zero.	

\begin{figure}[!ht]
\begin{minipage}{0.7\textwidth}
\subfigure[Solid lines: $G_{p_c}^D$ with $\alpha=0.05, \beta=0.02$. Dotted lines: $G_{p_c}$]{\includegraphics[width=\textwidth]{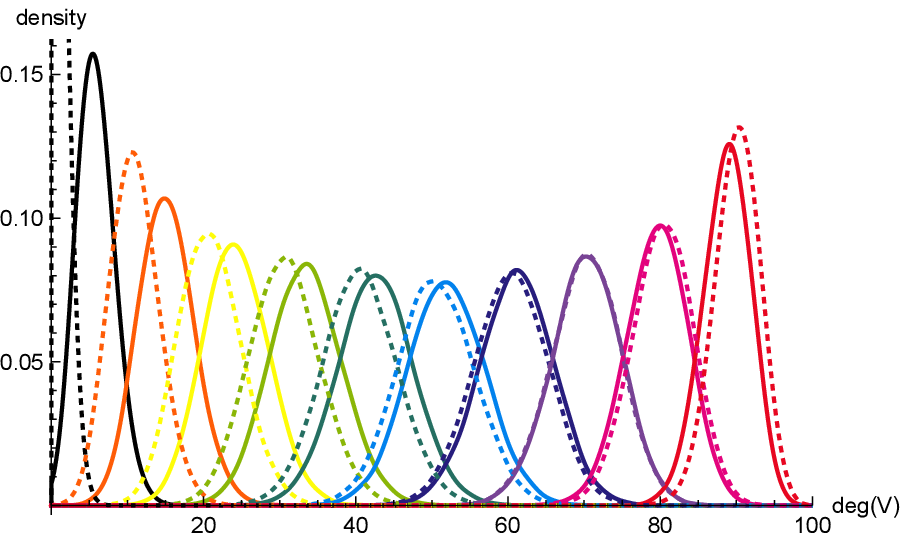}}
\subfigure[Solid lines: $G_{p_c}^D$ with $\alpha=0.02, \beta=0.05$. Dotted lines: $G_{p_c}$]{\includegraphics[width=\textwidth]{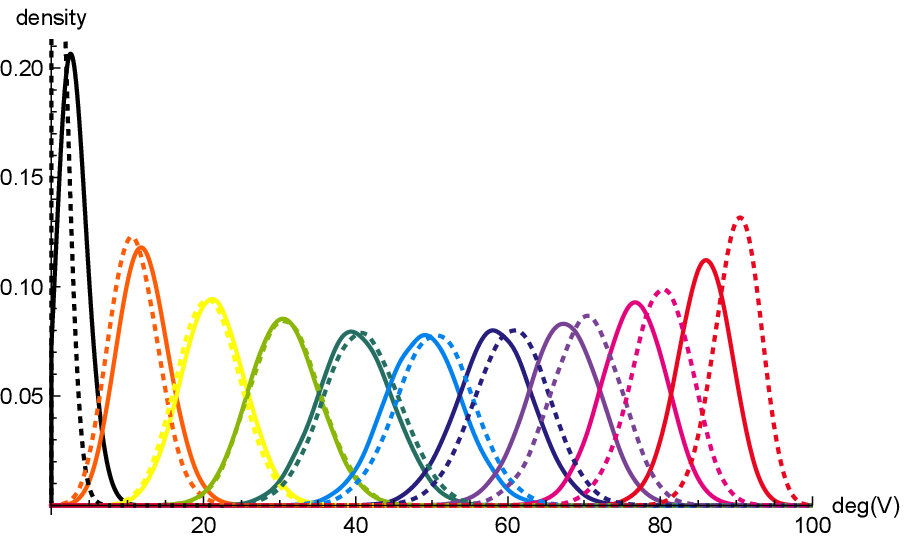}}
\end{minipage}
\quad
\begin{minipage}{0.2\textwidth}
\vspace*{-0.4cm}
\subfigure{\includegraphics[width=0.45\textwidth]{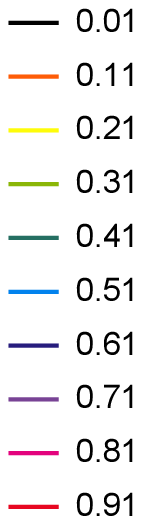}}

\vspace*{1.2cm}

\subfigure{\includegraphics[width=0.45\textwidth]{legend.eps}}
\end{minipage}
\caption{Densities of the node degree distributions for Erd{\H o}s-R{\'e}nyi networks of $n=100$ nodes and different parameters $p_c=0.01, ... , 0.91$ in steps of $0.1$ represented by colour. The densities of the node degree distributions for the respective original networks $G_{p_c}$ (dotted lines) and detected networks $G_{p_c}^D$ (solid lines) are shown.}
\label{degreedistrib}
\end{figure}

As an example of how the choice of $\alpha$ and consequently $\beta$ affects the estimated network characteristics, we consider Erd{\H o}s-Renyi networks $G_{p_c}$. 
The parameter $p_c$ was varied between $0.01$ and $0.99$ in steps of $0.01$. 
The detected networks $G^D_{p_c}$ were generated by artificially introducing false positive links with probability $\alpha$ and false negative links with probability $\beta$.
We varied $\alpha$ between $0.005$ and $0.1$ in steps of $0.001$, the relation between $\alpha$ and $\beta$ was fixed by
\begin{equation}\label{alphabeta}
\beta=\frac{10^{-3}}{\alpha},
\end{equation}
which represents a choice motivated by our simulations (Fig.~\ref{linear}).
Moreover, this choice corresponds to a method, which has high sensitivity and specificity, i.e., $0.005<\alpha,\beta<0.2$.
For each value of $p_c$ and $\alpha$, $200$ networks with $n = 100$ nodes were generated. 
Figure~\ref{degreedistrib} shows the true densities of the node degree derived from $G_{p_c}$ (dashed lines) together with the average densities derived from the detected networks $G^D_{p_c}$ (solid lines).
Results for two different values of $\alpha$ are shown.
Different colours represent different Erd{\H o}s-Renyi networks defined by the parameter $p_c$, for clarity, densities are plotted for $p_c$ in steps of $0.1$ only.

\begin{figure}[!ht]
\centering
\includegraphics[width=0.9\textwidth]{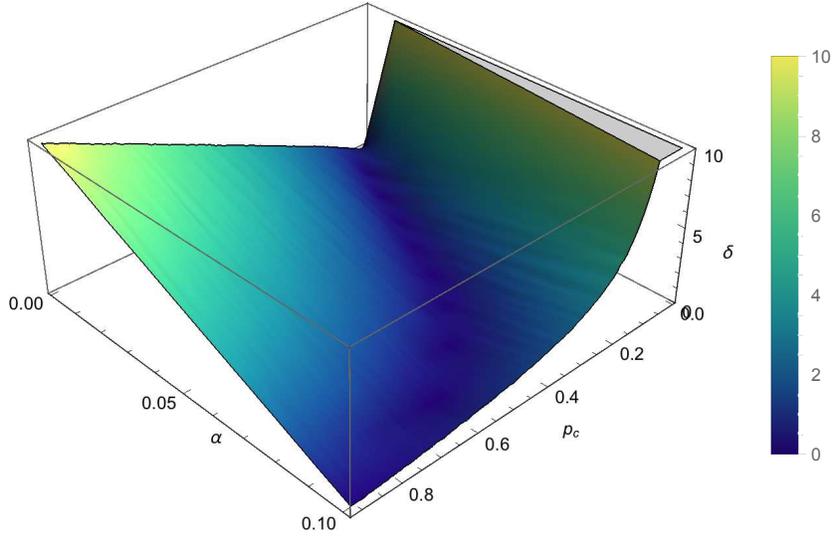}
\caption{Distance $\delta$ between node degree distributions of Erd{\H o}s-R{\'e}nyi networks with 100 nodes depending on $\alpha$ and $p_c$. Distance $\delta$ is measured by calculating the difference between the mean of two corresponding distributions, Eq.~(\ref{dist}). Colour code expresses distance values.}
\label{3Dvertex}
\end{figure}

The distances (Eq.~\ref{dist}) between the true density and the detected density for each pair of $p_c$ and $\alpha$ are shown in Fig.~\ref{3Dvertex}.
For some values of $p_c$ the distance is negligible, which means the detected node degree is almost identical to the true node degree.
The optimal $\alpha$, i.e. the one with the smallest distance between true density and detected density, depends on $p_c$. 

\begin{figure}[!ht]
\centering
\includegraphics[width=0.75\textwidth]{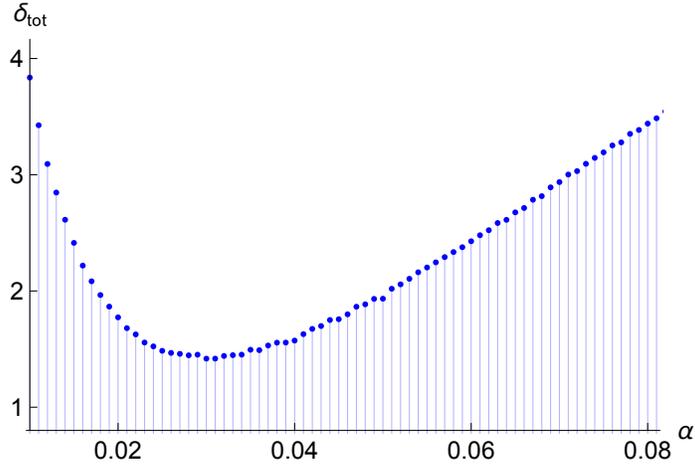}
\caption{Total distances $\delta_{tot}$ between node degree distributions of Erd{\H o}s-R{\'e}nyi networks depending on $\alpha$. The minimum is located at $\alpha = 0.030$.}
\label{integral}
\end{figure}

To have a general result for the optimal choice of $\alpha$ when estimating a network characteristic of a given network topology, we sum over $p_c$ to marginalise out the influence of $p_c$ for each $\alpha$. We call this integrated quantity the total distance $\delta_{tot}$, i.e. 
\begin{equation}
\delta_{tot}=\sum_{p_c}\delta(p_c).
\label{deltatot}
\end{equation}
To identify the optimal choice of $\alpha$, we are interested in finding where the minimum of the total distance $\delta_{tot}$ is located.
Figure~\ref{integral} shows $\delta_{tot}$ for the example of the node degree of Erd{\H o}s-R{\'e}nyi networks. 
In this example, the minimum of $\delta_{tot}$ is located at $\alpha=0.030$.
This suggests that in order to optimally reconstruct the node degree of an Erd{\H o}s-R{\'e}nyi network $\alpha=0.03$ should be chosen, which is close to the standard choice of $\alpha=0.05$ but distinctively smaller.

\section{Results}
\label{Results}

\begin{figure}[!t]
\centering
\begin{tabular}{m{0.08\textwidth} m{0.25\textwidth}m{0.25\textwidth}m{0.25\textwidth}| m{0.05\textwidth}}
& Erd{\H o}s-R{\'e}nyi networks & Barab{\'a}si-Albert networks & Watts-Strogatz networks&\\
node degree & 
\includegraphics[width=0.25\textwidth]{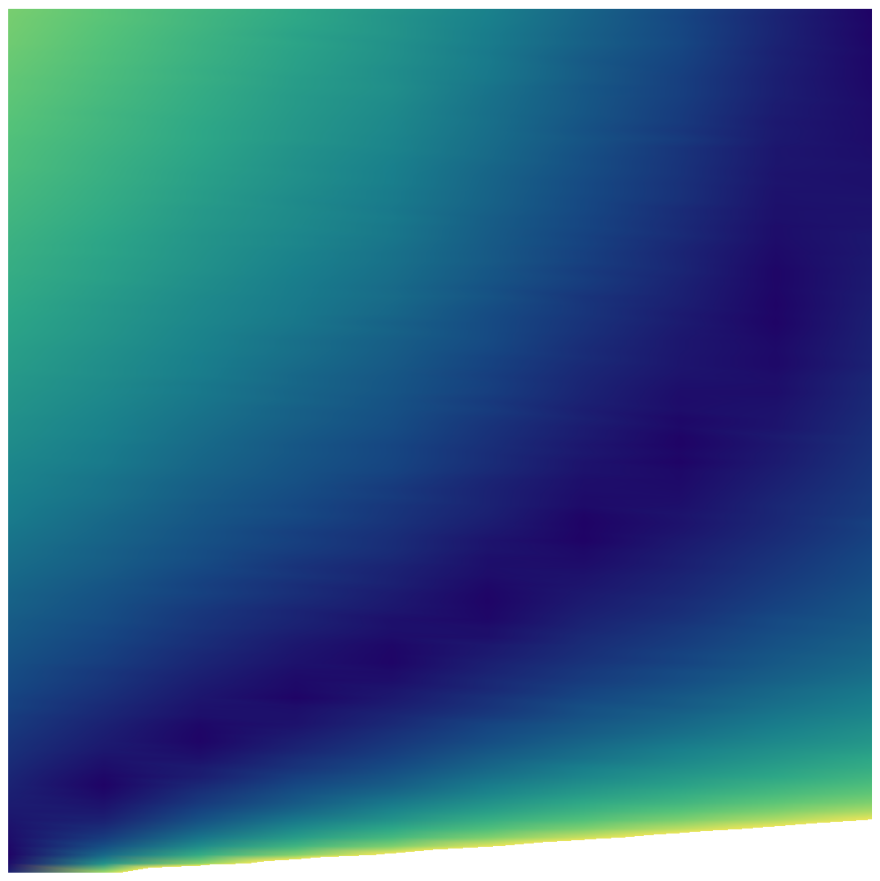} & 
\includegraphics[width=0.25\textwidth]{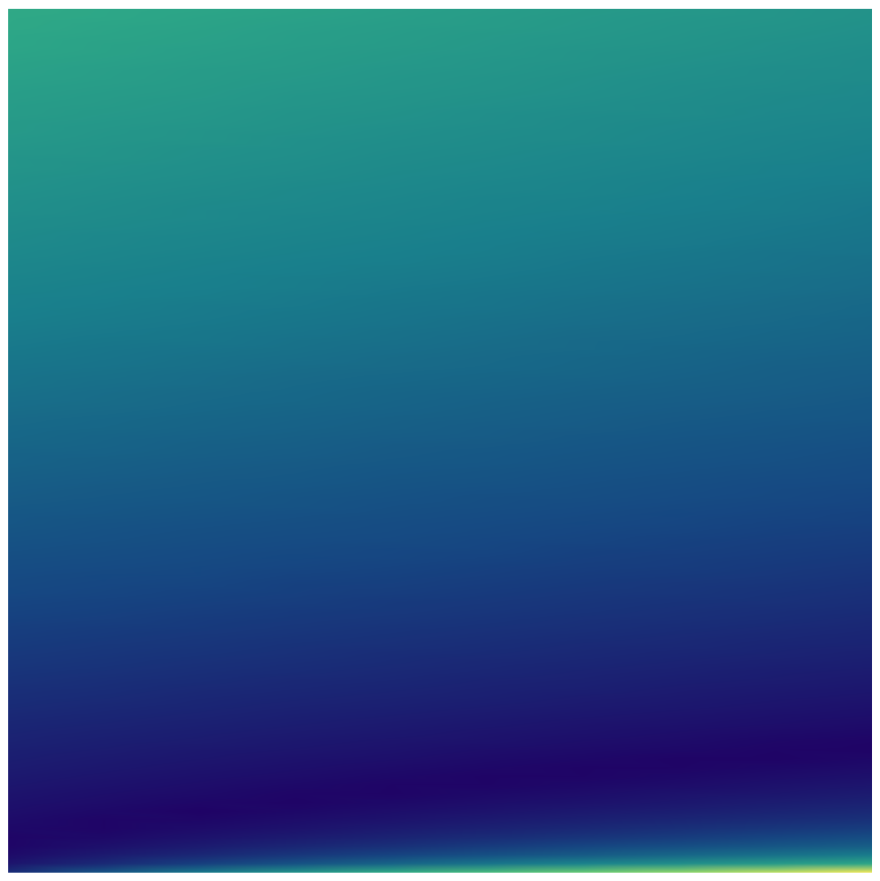} &
\includegraphics[width=0.25\textwidth]{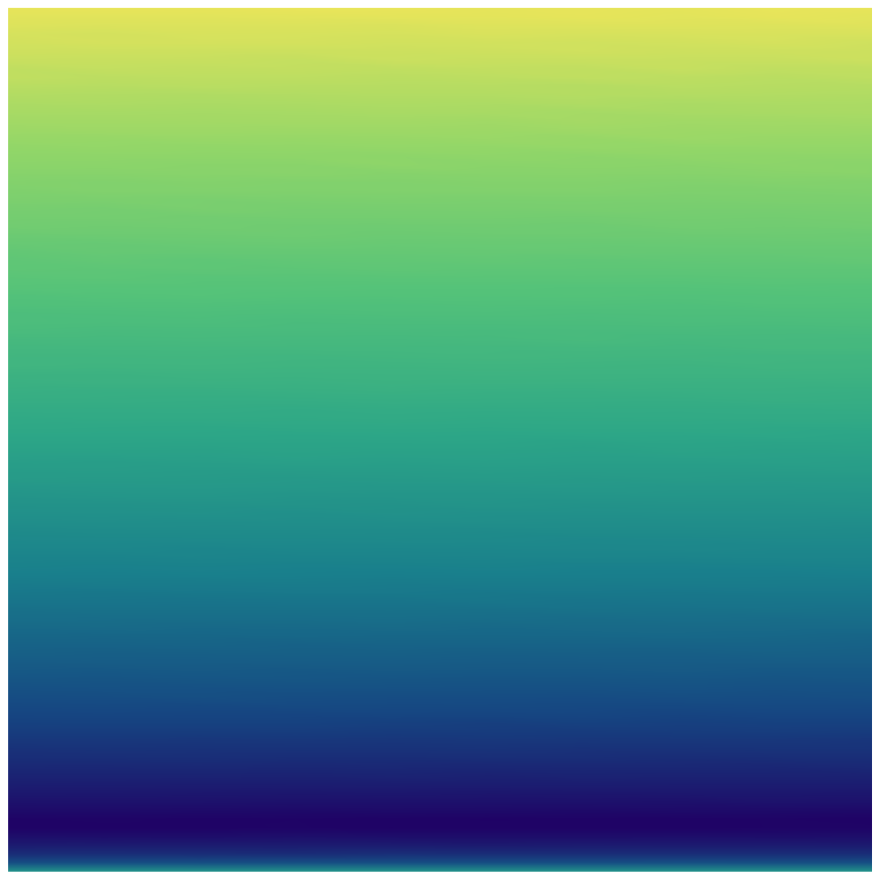}&$\alpha$\\
number of links &
\includegraphics[width=0.25\textwidth]{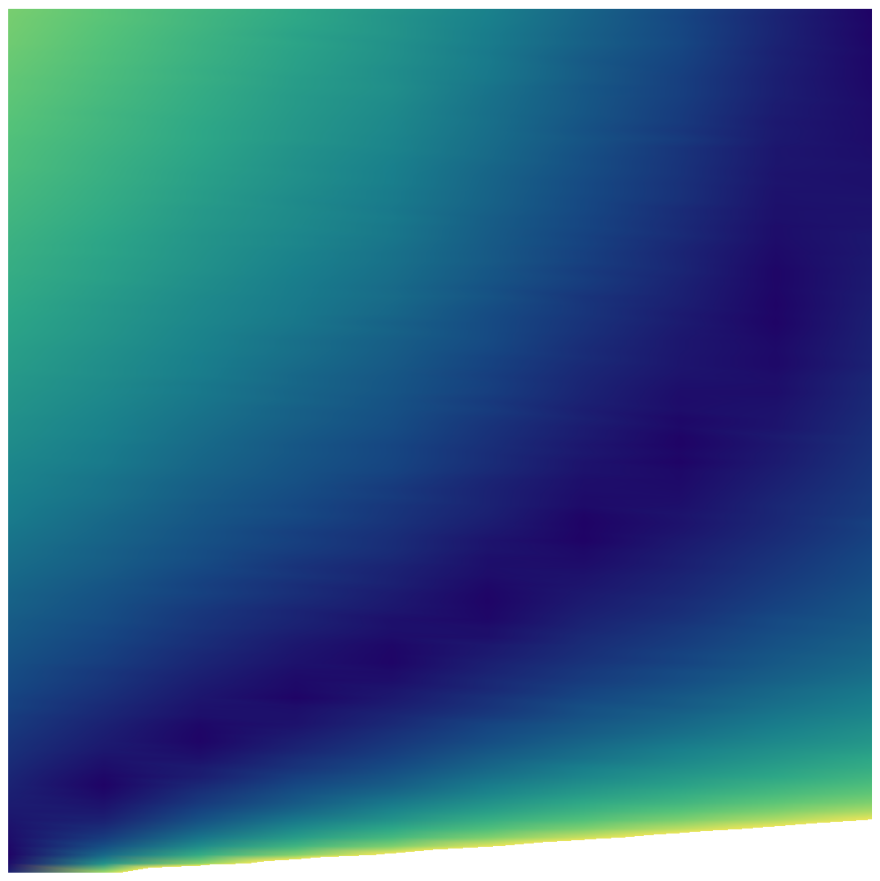}&
\includegraphics[width=0.25\textwidth]{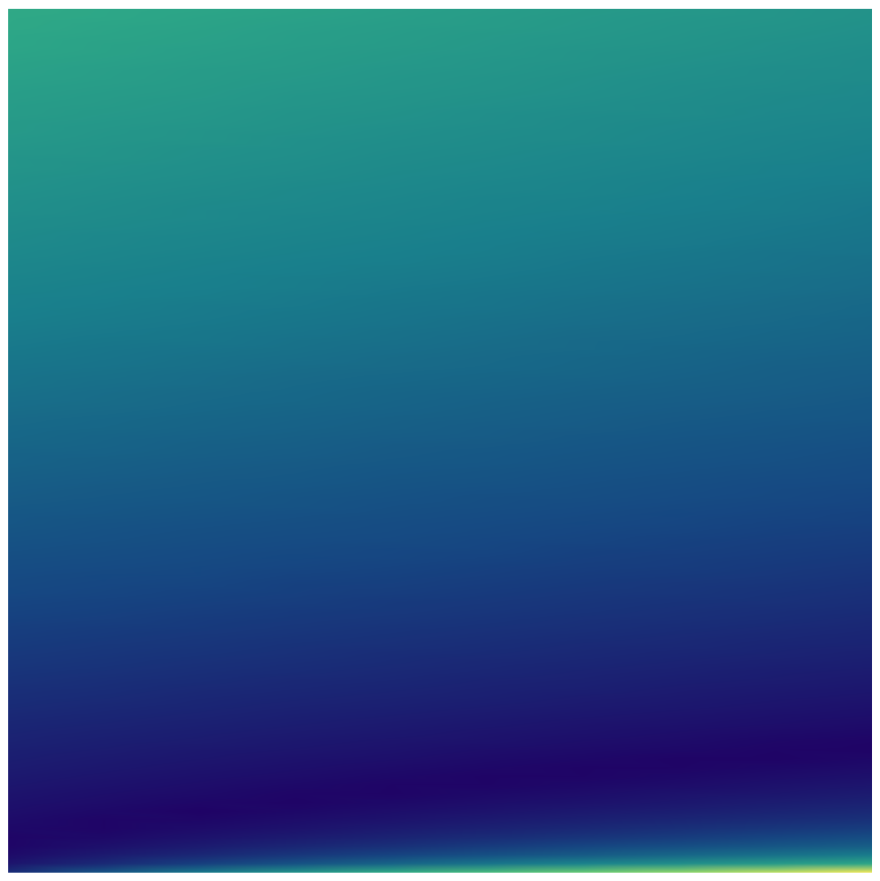}&
\includegraphics[width=0.25\textwidth]{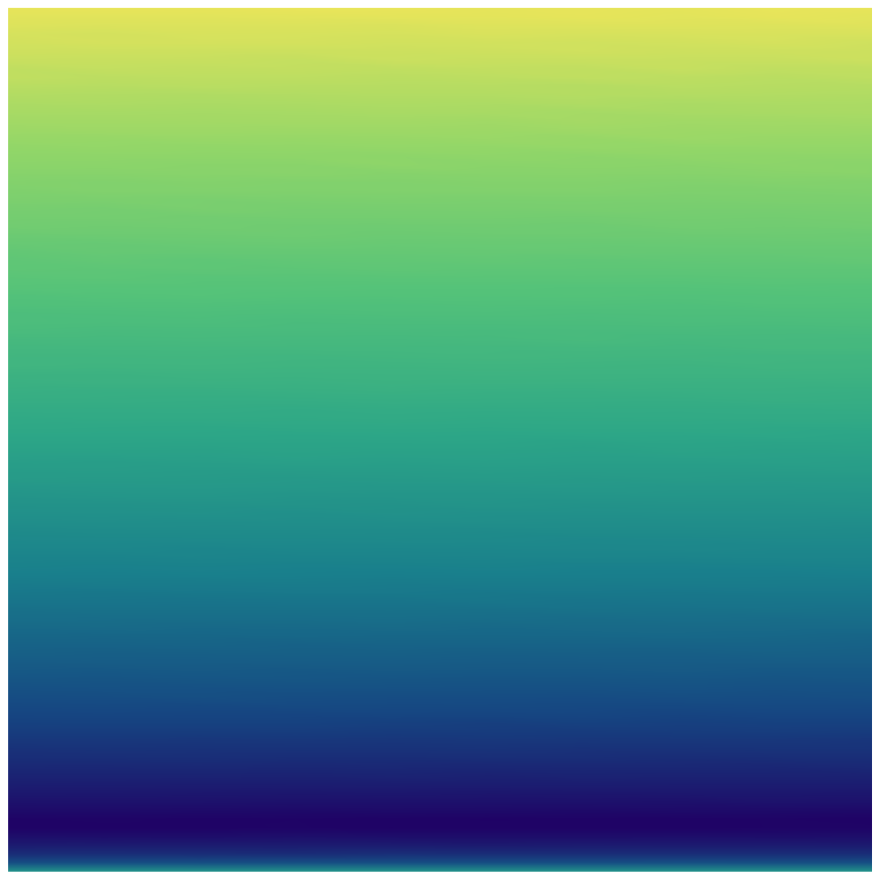}&$\alpha$\\
efficiency&
\includegraphics[width=0.25\textwidth]{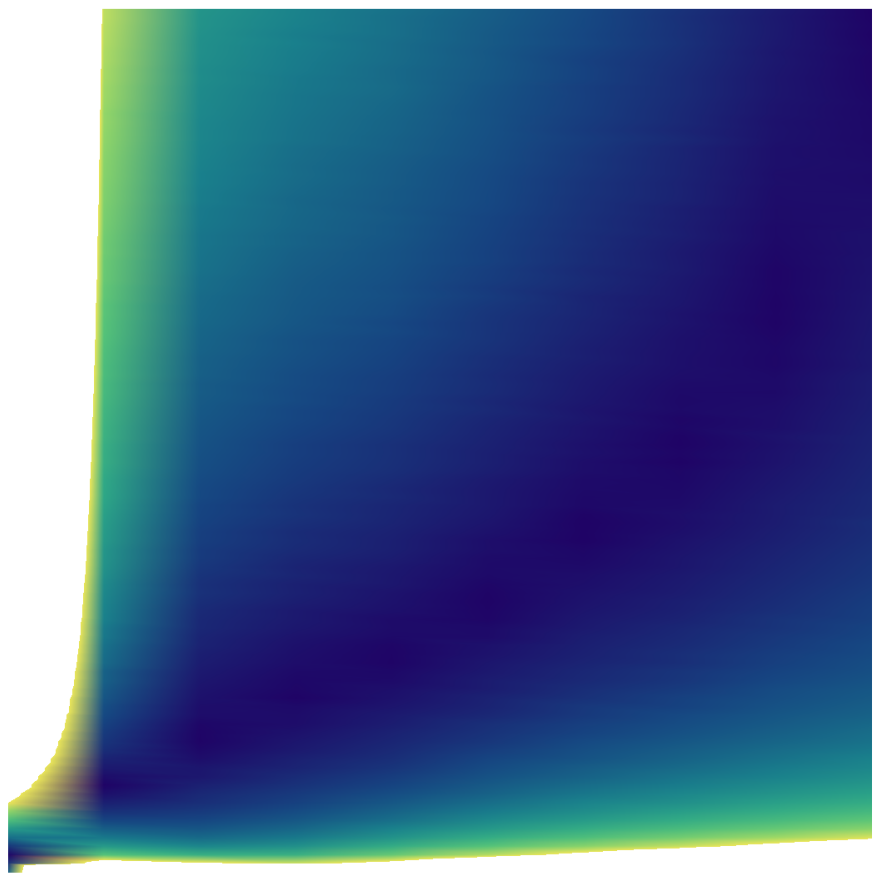}&
\includegraphics[width=0.25\textwidth]{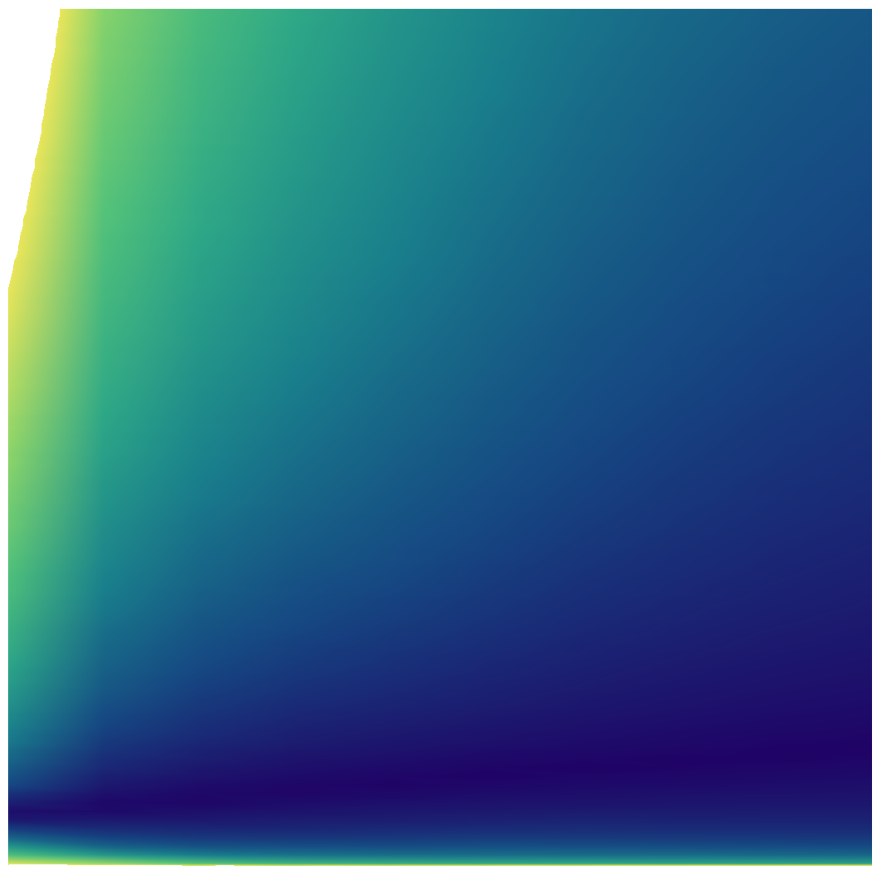}&
\includegraphics[width=0.25\textwidth]{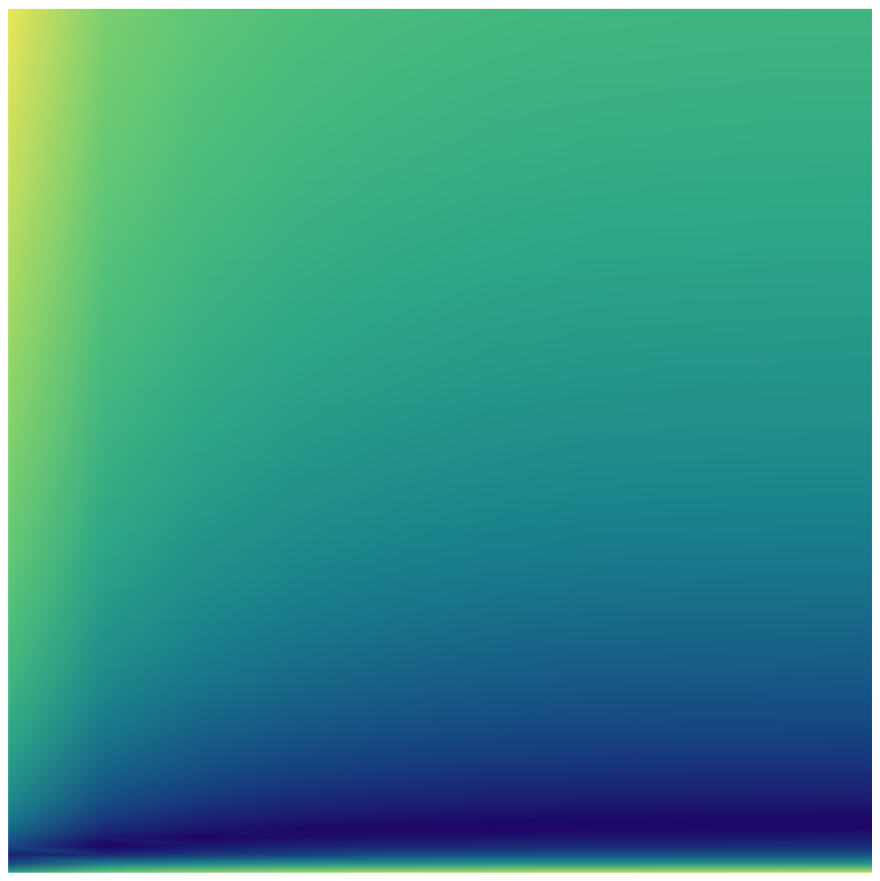}&$\alpha$\\
global clustering coefficient &
\includegraphics[width=0.25\textwidth]{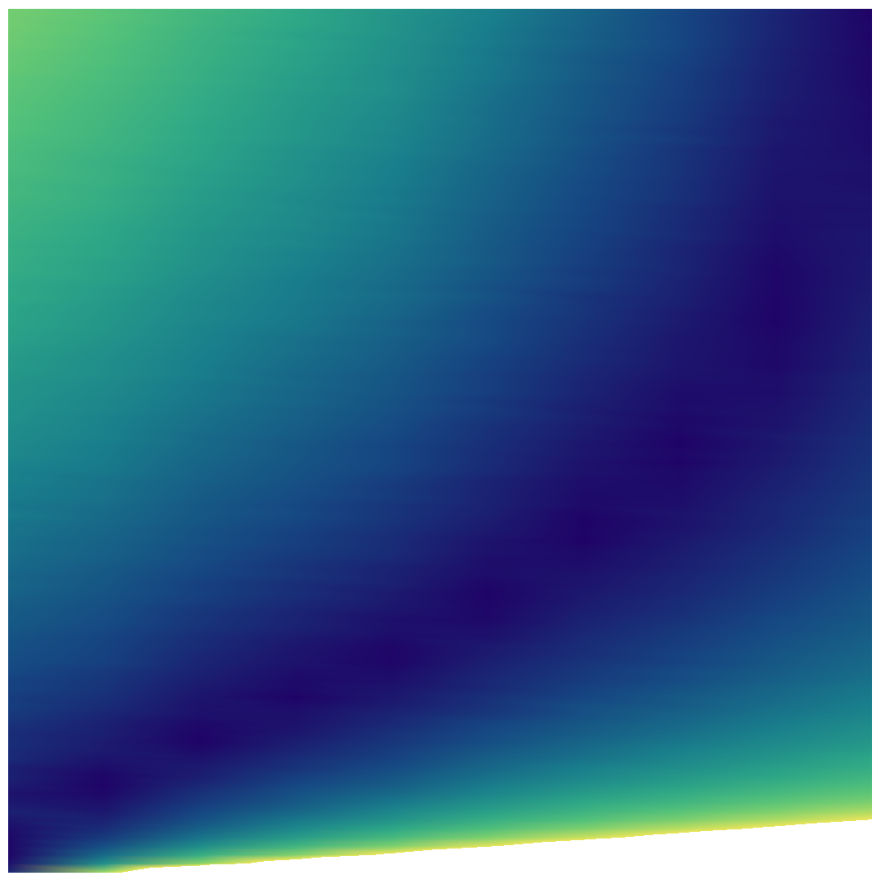}&
\includegraphics[width=0.25\textwidth]{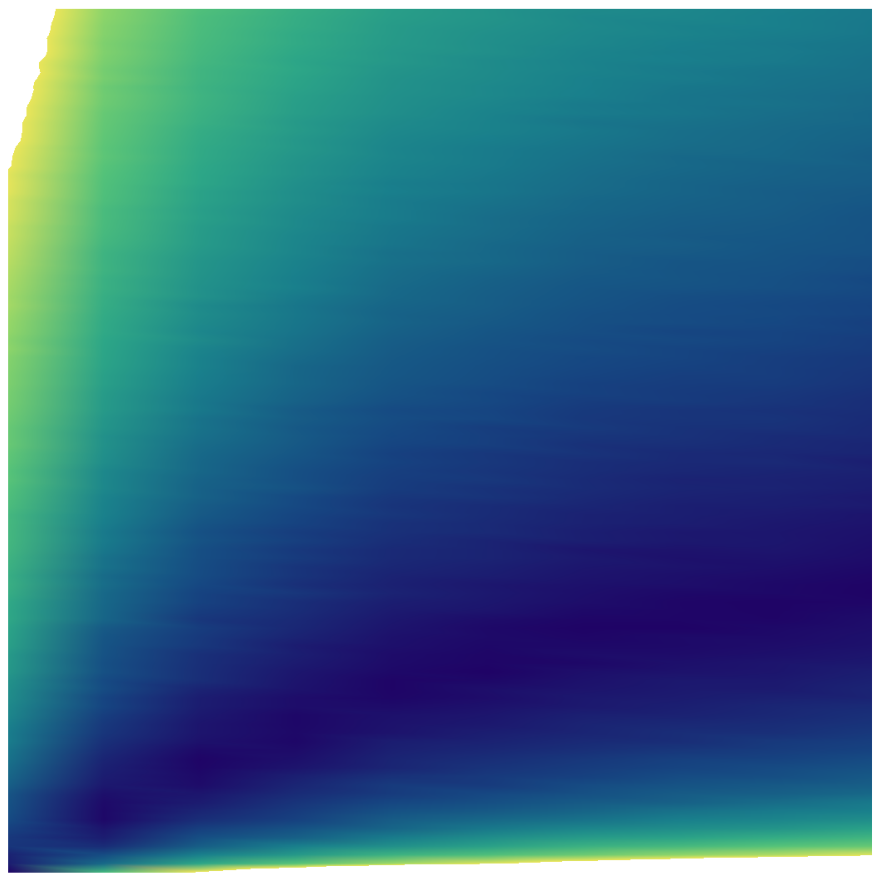}&
\includegraphics[width=0.25\textwidth]{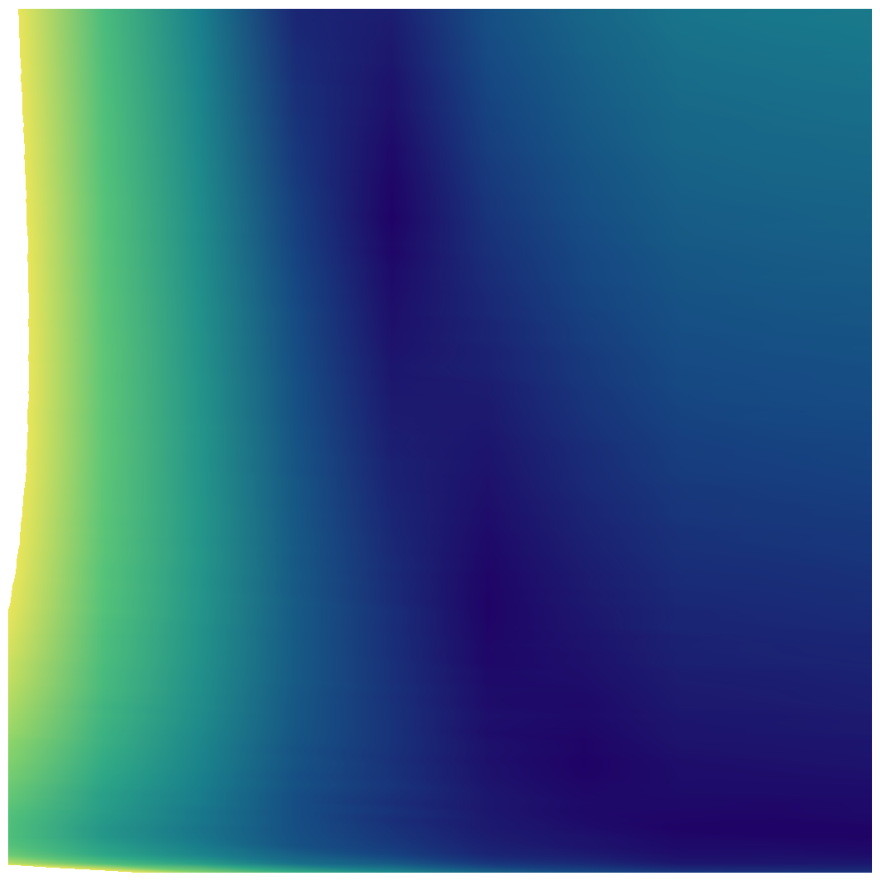}&$\alpha$\\
\hline
&$\qquad\qquad p_c$&$\qquad\qquad b$&$\qquad\qquad p_r$&\\
\end{tabular}
\caption{Distance $\delta$ for Erd{\H o}s-R{\'e}nyi, Barab{\'a}si-Albert, and Watts-Strogatz networks with 100 nodes. Distance $\delta$ is measured by calculating the difference between the mean of two corresponding distributions, Eq.~(\ref{dist}), and it is expressed by colour code, from blue to yellow. For each network topology we vary the control parameter $p_c$ from $0.01$ to $0.99$ in steps of $0.01$, $b$ from $1$ to $10$ in steps of $1$, or $p_r$ from $0.01$ to $0.99$ in steps of $0.01$ on the $x-$axis, and the probability of false positive $\alpha$ from $0.005$ to $0.1$ in steps of $0.001$ on the $y-$axis.}
\label{3Dgeneral}
\end{figure}

We applied our analysis to Erd{\H o}s-R{\'e}nyi, Watts-Strogatz and Barab{\'a}si-Albert networks.
For each network topology, we investigated four different network characteristics: node degree, number of links, global clustering coefficient and efficiency.
The distance $\delta$, which depends on both $\alpha$ and the parameter of the network topology ($p_c,b$ or $p_r$) is presented as density plot for all the investigated characteristics and network topologies in Fig.~\ref{3Dgeneral}.
All 12 investigated scenarios show a dependence of the distance on the choice of $\alpha$, suggesting that an optimum exists.
For some scenarios, in particular node degree and number of links for Watts-Strogatz and Barab{\'a}si-Albert networks, dependence of the distance on the parameter ($p_r$ or $b$) is negligible.
For other scenarios such as the global clustering coefficient in Watts-Strogatz networks the question arises if marginalising out the influence of $p_r$ is distorting the results.
Detailed results for each network topology are presented below.

For Erd{\H o}s-R{\'e}nyi networks of $n=50$, $n=100$, and $n=250$ nodes we varied $p_c$ from $0.01$ to $0.99$ in steps of $0.01$.
Figure \ref{degreedistrib} shows an example of some of these values in steps of $0.1$.
The results of the total distance $\delta_{tot}$ for the node degree of Erd{\H o}s-R{\'e}nyi networks are shown in Fig.~\ref{integral}.
The minimum of $\delta_{tot}$ is located at $\alpha = 0.030$ ($\beta = 0.033$).
For the remaining network characteristics, the Erd{\H o}s-R{\'e}nyi networks also show a clear minimum of the total distance in dependence on $\alpha$.
The specific values of $\alpha$ for the respective minimal total distances however vary; they are summarised in Table \ref{table}.
The optimal $\alpha$ for efficiency is noticeably smaller than for the other network characteristics.
Moreover, we chose a broad range for $p_c$ to cover the broad spectrum of Erd{\H o}s-R{\'e}nyi networks.
Marginalising out the dependence of the distance $\delta$ on $p_c$ may therefore be distorting our results (see also dependence on $p_c$ in Fig.~\ref{3Dgeneral}).
For a specific application we thus recommend narrowing the range of $p_c$ to values relevant for the application.

\begin{table}[!ht]
\begin{center}
\begin{tabular}{| l | c | c | c | c | c | c |}
\hline
Network Topology  &\multicolumn{2}{|c|}{$n=50$}&\multicolumn{2}{|c|}{$n=100$}&\multicolumn{2}{|c|}{$n=250$}\\
\qquad and Characteristic & $\alpha$ & $\beta$ &$\alpha$ & $\beta$&$\alpha$ & $\beta$\\
\hline
Erd{\H o}s-R{\'e}nyi: &&&&&&\\
\qquad node degree             &0.031&0.032&0.030&0.033&0.031&0.032\\
\qquad number of links           &0.031&0.032&0.030&0.033&0.031&0.032\\
\qquad global clustering coeff &0.035&0.029&0.031&0.032&0.031&0.032\\
\qquad efficiency                   &0.016&0.063&0.012&0.083&0.020&0.050\\
\hline
Barab{\'a}si-Albert: &&&&&&\\
\qquad node degree             &0.018&0.056& 0.012 & 0.083 & 0.007&0.143\\
\qquad number of links           &0.018&0.056& 0.012 & 0.083 & 0.007&0.143\\
\qquad global clustering coeff &0.024&0.042& 0.021 & 0.048 & 0.019&0.053\\
\qquad efficiency                   &0.015&0.067& 0.010 & 0.100 & 0.007&0.143\\
\hline
Watts-Strogatz: &&&&&&\\
\qquad node degree             &0.009&0.111& 0.007 & 0.143 & 0.004&0.250\\
\qquad number of links           &0.009&0.111& 0.007 & 0.143 & 0.004&0.250\\
\qquad global clustering coeff &0.008&0.125& 0.006 & 0.167 & 0.004&0.250\\
\qquad efficiency                   &0.009&0.111& 0.006 & 0.167 & 0.004&0.250\\
\hline
\end{tabular}
\end{center}
\caption{Table of $\alpha$ and $\beta$ values for minimal total distances $\delta_{tot}$ of each network topology and characteristic.}
\label{table}
\end{table}

\begin{figure}[!ht]
\centering
\includegraphics[width=0.75\textwidth]{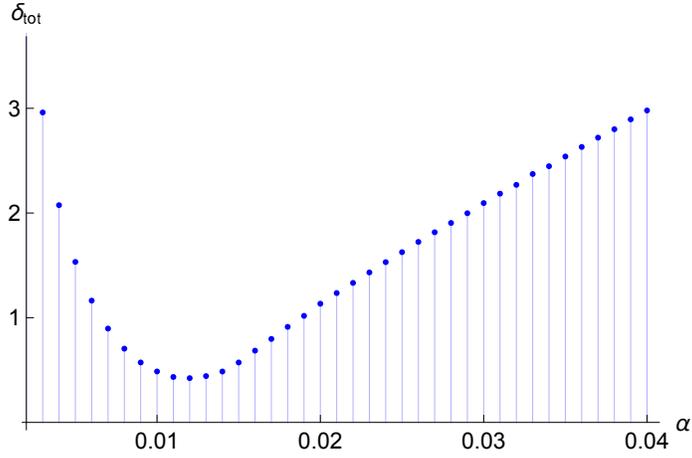}
\caption{Total distance $\delta_{tot}$ between node degree distributions of Barab{\'a}si-Albert networks of $n=100$ nodes depending on $\alpha$. The minimum is located at  $\alpha = 0.012$ ($\beta = 0.083$).}
\label{integral2}
\end{figure}

The set of Barab{\'a}si-Albert networks of $n=50$, $n=100$, and $n=250$ nodes was chosen with average degree typical for networks in neuroscience \citep{Papo2014,Stanley2013} by using parameters $b=1$ to $b=10$ varying in steps of $1$.
The total distances $\delta_{tot}$ for the node degree of networks with $n = 100$ nodes are shown in Fig.~\ref{integral2}.
The minimum is found for $\alpha = 0.012$ ($\beta = 0.083$), it is more pronounced than that for the Erd{\H o}s-R{\'e}nyi networks.
Again, the other network characteristics and number of nodes all show a single minimum. 
The values for optimal $\alpha$ and $\beta$ are summarised in Table \ref{table}.
For this network topology a noticeably different optimal value for $\alpha$ was found for the clustering coefficient.

\begin{figure}[!ht]
\centering
\includegraphics[width=0.75\textwidth]{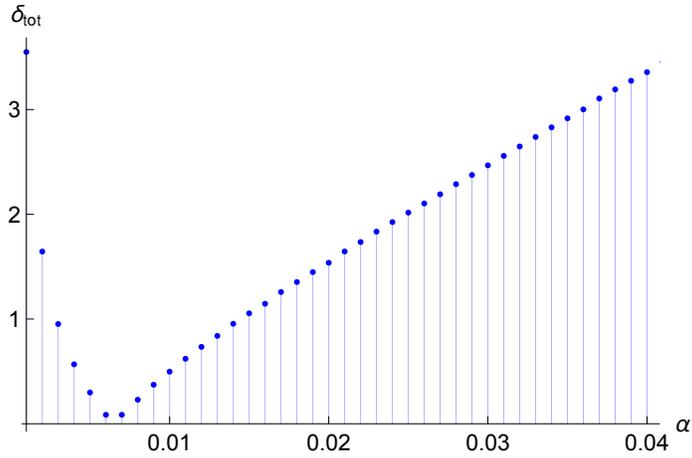}
\caption{Total distance $\delta_{tot}$ between node degree distributions for Watts-Strogatz networks of $n=100$ nodes depending on $\alpha$. The minimum is located at  $\alpha = 0.007$ ($\beta = 0.143$).}
\label{integral3}
\end{figure}

Finally, we considered a set of Watts-Strogatz networks of $n=50$, $n=100$, and $n=250$ nodes with parameter $p_r$ varied between $0.01$ and $0.99$ in steps of $0.01$.
We analysed distances between the distributions of the node degree, the number of links, the global clustering coefficient and the efficiency.
The minimum of the total distance for the node degree of networks with $n=100$ nodes is found for $\alpha = 0.007$ ($\beta = 0.143$).
The total distances between node degree distributions for these networks are shown in Fig.~\ref{integral3}.
For all four characteristics and different values of $n$ clear minima can be identified and the values for optimal $\alpha$ are similar (Table~\ref{table}).
The results for the efficiency however have to be interpreted with case as the distance showed a clear dependence on the parameter $p_r$ (see Fig.~\ref{3Dgeneral}).

\section{Discussion}
\label{Discussion}

We consider three topologies of networks Erd{\H o}s-R{\'e}nyi, Watts-Strogatz, and Barab{\'a}si-Albert.
For each topology, and for a specific characteristic, e.g. the node degree distribution, the number of links, the efficiency or global clustering coefficient, the rate of false positive and false negative conclusions about links can be optimally chosen in order to have less biased reconstruction.

For Erd{\H o}s-R{\'e}nyi networks, the values for $\alpha$ identified with our method are close to standard choice of $\alpha$ of $0.05$.
Standard alpha values are suboptimal when the topology of network is different.
For the set of Barab{\'a}si-Albert networks we found that the value for $\alpha = 0.012$ and consequently $\beta = 0.083$ yields the most reliable results of the node degree.
In this case, standard alpha values lead to a bigger distance between distributions of the node degree.
The Watts-Strogatz networks yield the most reliable results for an even smaller value for $\alpha = 0.007$ and consequently $\beta = 0.154$. 
Moreover, for the optimal choice of $\alpha$ the corresponding $\beta$ is rather high.
This shows that the reliability of detecting individual false negative links in a network is less important than failing to recognise false positive links when network characteristics are estimated.
Accepting a high rate of false negative links may thus be required when the aim is to infer a specific network characteristic.

This work shows that the standard choice of $\alpha$ of $0.05$ is not optimal when the aim is to reconstruct the entire network topology.
Moreover, $\alpha$ needs to be adjusted depending on specific network topologies and characteristics.
For example, consider Erd{\H o}s-R{\'e}nyi networks with $p_c=0.11$ and assume the relationship between $\alpha$ and $\beta$ is Eq.~\ref{alphabeta}.
As result of 200 simulations, the mean of the node degree distribution of the original network $G_{p_c}$ is 11 and the mean for estimation using $\alpha=0.05$ is 15.
Choosing $\alpha=0.03$ results in a mean of the node degree distribution of 13.
The choice of $\alpha=0.03$ is motivated by the assumption that the original network is known to be an Erd{\H o}s-R{\'e}nyi network with unknown parameter $p_c$, see Table~\ref{table}.
For the same study, when the aim is to infer the efficiency $\epsilon$, we calculate $\epsilon=0.51$ for the true network, $\epsilon=0.56$ for the one with $\alpha=0.05$, and $\epsilon=0.51$ when $\alpha=0.012$.
The more we know about the network we want to infer the more accurate the reconstruction is since the simulation study can be tuned accordingly.

As mentioned in Section \ref{InferenceReliability}, the relationship between $\alpha$ and $\beta$ depends on the number of data points $N$, therefore the values of $\alpha$ and $\beta$ leading to the minimal distance will change for different values of $N$.
Nevertheless, the results will remain qualitatively the same.

The size of the network, i.e. the number of nodes, also influences the result.
The number of false positive and false negative conclusions about the presence of links depends on the number of total links in the network.
Keeping the same rate of $\alpha$ and $\beta$ and increasing, for example, the size of the network, leads to larger number of false positive and false negative detections of links.
As shown in Table~\ref{table}, for Erd{\H o}s-R{\'e}nyi networks the values of $\alpha$ and $\beta$ leading to the minimal distance almost do not change.
The reason is that the number of links increases proportionally with the number of nodes for each $p_c$.
This does not happen for Barab{\'a}si-Albert and Watts-Strogatz networks; the values of $\alpha$ leading to the minimal distance present a decreasing trend because of their constructions.

We considered the node degree distribution, the number of links, the efficiency, and the global clustering coefficient as example characteristics to show that the results depend on the characteristic under investigation. 
Nevertheless, our approach can be readily applied to other characteristics, as well as other network topologies.

\section{Conclusion}
\label{Conclusion}

False conclusions about the presence of links in a network typically alter network characteristics, such as the node degree distribution, the number of links, the global clustering coefficient and the efficiency.
Identification of the underlying network topology relies on these characteristics and is thus hindered by false conclusions about links as well.
For these reasons, the analysis of false positive and false negative conclusions about links is of key importance.

In this manuscript, assuming to know the underlying network topology, we investigate the influence of false positive and false negative conclusions about links in a network.
We show that the values of $\alpha$ and $\beta$ leading to minimal distance (difference in mean values) between the true network and the biased one change depending not only on the network topology, but also on the network characteristic of interest.
Therefore, in the {\em Inverse Problem}, when the challenge is to infer the network topology from data, different values for $\alpha$ and $\beta$ might be favourable when estimating different characteristics.
We speculate that our simulation study can be used as an iterative procedure to achieve a better network reconstruction. 
Namely, when the network topology is not known a priori, various values for $\alpha$ can be chosen to perform the first iteration step of the network reconstruction. 
The result of this first step gives an idea of the network topology we want to infer. 
For the second iteration step the value for $\alpha$ can be adjusted according to the findings of the first step. 
This procedure can be iterated using the simulation study that we suggest in this paper in each iteration step, ultimately leading to a reconstruction of the network tailored to its previously unknown network topology.

This result suggests that in the Neurosciences, as well as in other scientific fields, various values for statistical inference could be considered within a simulation study to determine the optimal $\alpha$ for the network characteristic of interest.
If several network characteristics are of interest, it may be useful to adjust the value of $\alpha$ for each characteristic.

\section*{Acknowledgements}

This project has received funding from the European Union's Horizon 2020 research and innovation programme under the Marie Sklodowska-Curie grant agreement No 642563.

The authors declare no competing financial interests.

\bibliographystyle{elsarticle-harv}

\bibliography{wns-bib}

\end{document}